# HEAT: History-Enhanced Dual-phase Actor-Critic Algorithm with A Shared Transformer

Hong Yang
*College of Computer and Data Science*
*Fuzhou University*
Fuzhou, China
hongyang0066@gmail.com

*Abstract*—For a single-gateway LoRaWAN network, this study proposed a history-enhanced two-phase actor-critic algorithm with a shared transformer algorithm (HEAT) to improve network performance. HEAT considers uplink parameters and often neglected downlink parameters, and effectively integrates offline and online reinforcement learning, using historical data and real-time interaction to improve model performance. In addition, this study developed an open source LoRaWAN network simulator LoRaWANSim. The simulator considers the demodulator lock effect and supports multi-channel, multi-demodulator and bidirectional communication. Simulation experiments show that compared with the best results of all compared algorithms, HEAT improves the packet success rate and energy efficiency by 15% and 95%, respectively.

*Keywords—Internet of Things, LoRaWAN, Communications and Networking, Deep Reinforcement Learning, Offline Reinforcement Learning*

## 1 Introduction

Although the LoRaWAN network can support a larger node scale than the LoRa private network, as the number of devices increases, the performance of the LoRaWAN network in terms of network congestion and energy consumption faces significant challenges. The limited spectrum resources and channel congestion will lead to a decrease in the communication efficiency of the network, which in turn affects the reliability of data transmission. How to achieve efficient and energy-saving resource allocation while ensuring network performance remains a key issue. In order to improve the overall performance of the LoRaWAN network, optimizing the transmission strategy parameters such as the spreading factor, transmit power, and receive window of the uplink and downlink is considered to be an effective means. By reasonably configuring these parameters, network conflicts can be effectively reduced, signal attenuation can be reduced, and signal coverage can be increased, thereby improving network reliability and communication quality. However, most of the existing optimization methods focus on the adjustment of the spreading factor and transmit power of the uplink, and rarely consider the impact of the downlink on network performance.

To address this problem, this chapter proposes a History-Enhanced two-phase Actor-Critic algorithm with a shared Transformer (HEAT), which aims to improve the resource allocation strategy of the LoRaWAN network and improve the overall performance of the network. This chapter conducts multiple sets of comparative experiments between HEAT and various popular methods under different device densities and traffic intensities to verify the effectiveness of HEAT.

# 2 System Model and Problem Representation

In order to efficiently verify the effectiveness of various LoRaWAN resource allocation strategies, this section describes and models the LoRa link behavior and the LoRaWAN standard in detail. Subsequently, this section proposes the target problem of LoRaWAN resource allocation and expresses the target problem as a Markov decision process.

## 2.1 LoRaWAN Network Model

This paper considers a LoRaWAN network that supports uplink and downlink communications. The gateways and terminal nodes are distributed in the deployment area $\mathcal{V} \subseteq \mathbb{R}^2$, where a single gateway is located at the origin of the coordinate system, and the terminal nodes are randomly distributed around the gateway. The deployment area $\mathcal{V}$ is a disk with a radius of $R$ and an area of $S = |\mathcal{V}| = \pi R^2$. The terminal nodes are randomly distributed around the gateway following a homogeneous Poisson point process $\Phi$ with an intensity function of $\lambda$. Assuming that the area $\mathcal{V}$ contains $N$ terminal nodes, then $\lambda = S/N$. Each terminal node generates $\delta$ packets per minute, and the timing of packet generation follows a Poisson distribution [1] to reflect the discreteness and randomness of transmission in LPWAN. The received signal strength $RSSI$ depends on the transmit power $P$ and the gain and loss on the communication path, expressed as:

$$RSSI_i = P_i G_i L_i \zeta_i. \tag{1}$$

Where $\mathcal{D}$ represents the set of all terminal nodes, and the subscript $i \in \mathcal{D}$ represents a specific terminal node. $G_i$ represents the combined gain of the transmitting antenna and the receiving antenna. $L_i = (c/4\pi f d_i)^2$ is the Friss path loss, where $f$ and $c$ are the carrier frequency and the speed of light, respectively. $d_i$ is the Euclidean distance between device $i$ and the gateway. In order to more accurately simulate and predict the propagation characteristics of wireless signals in actual environments, the small-scale Rayleigh fading term $\zeta_i \sim \mathrm{EXP}(1)$ is introduced in this model to describe the rapid changes caused by reflection, refraction and scattering of signals encountered during propagation due to obstacles. If the received signal strength $RSSI$ is higher than the sensitivity threshold $\tau_s$ corresponding to the spreading factor $SF = s$, the demodulator can successfully capture the LoRa signal. Table 1 shows the characteristics of SF.

Table 1 LoRaWAN Spreading Factor Characteristics

| SF | $\tau_s$ (dBm) | $\tau_d$ (dB) | Data Rate (bit/sec) |
|---|---|---|---|
| 7 | -127 | -7.5 | 5469 |
| 8 | -129 | -10 | 3125 |
| 9 | -132.5 | -12.5 | 1758 |
| 10 | -135.5 | -15 | 977 |
| 11 | -138 | -17.5 | 537 |
| 12 | -141 | -20 | 293 |

In the LoRaWAN network, since the terminal nodes randomly send data packets using the ALOHA protocol and perform frequency hopping for each transmission, the terminal nodes are very susceptible to interference from other terminal nodes during the transmission process. Considering interference and noise, the signal-to-interference-plus-noise ratio (SINR) of the terminal node at the receiving end is expressed as:

$$SINR_i = \frac{RSSI_i}{\sigma^2 + \mathcal{J}_i}. \tag{2}$$

Among them, $\sigma^2$ represents the variance of Additive White Gaussian Noise (AWGN), defined as $\sigma^2 = N_0 + NF + 10\log(BW)\,[\mathrm{dBm}]$, $N_0$ is the noise power density, usually taken as $-174\,\mathrm{dBm/Hz}$. $NF$ is the receiver noise factor, which is related to the receiver hardware and is usually taken as 6dB. The interference term $\mathcal{J}_i$ represents the interference from other nodes to the terminal node $i$. If $SINR_i$ is higher than the demodulation threshold $\tau_d$ of the corresponding SF and no signal conflict occurs, the demodulator can successfully demodulate the LoRa signal.

At present, most related research works have simplified $\mathcal{J}_i$, assuming that there is no interference between EDs using different SFs, that is, different SFs are completely orthogonal. However, Ta et al. [2] and Bouazizi et al. [3] pointed out that in actual wireless environments, the impact of interference between different SFs on link performance cannot be ignored. In particular, when the interference source is closer to the gateway than the target terminal node and the network scale is large, this interference will seriously affect the network performance. Therefore, this study considers the incomplete orthogonality between different SFs and expresses its interference term as:

$$\mathcal{J}_i = \sum_{q=7}^{12} \sum_{j \in \mathcal{D}_q} \mathbb{I}_t(i,j)\mathbb{I}_c(i,j) RSSI_j \beta_{s_i,s_j}. \tag{3}$$

The set $\mathcal{D}_q$ represents the set of terminal nodes using SF $q$. The SF orthogonality coefficient $\beta_{s_i,s_j}$ is used to describe the orthogonality between the SFs used by terminal node $i$ and terminal node $j$. The specific values are shown in Table 2. The indicator functions $\mathbb{I}_t$ and $\mathbb{I}_c$ are used to describe whether different EDs send messages at the same time and whether they use the same channel, respectively.

Table 2 LoRa spreading factor orthogonality coefficient [3]

| $s_i$ \ $s_j$ | 7 | 8 | 9 | 10 | 11 | 12 |
|---|---|---|---|---|---|---|
| 7 | 1 | 0.104 | 0.062 | 0.041 | 0.029 | 0.021 |
| 8 | 0.104 | 1 | 0.073 | 0.043 | 0.029 | 0.020 |
| 9 | 0.062 | 0.073 | 1 | 0.052 | 0.030 | 0.020 |
| 10 | 0.041 | 0.043 | 0.052 | 1 | 0.037 | 0.021 |
| 11 | 0.029 | 0.029 | 0.030 | 0.037 | 1 | 0.026 |
| 12 | 0.021 | 0.020 | 0.020 | 0.021 | 0.026 | 1 |

In addition, this model also takes into account the capture effect in the LoRa demodulation process. The capture effect refers to the situation when multiple signals arrive at the receiver at the same time and frequency. If the power of one of the signals is at least a certain threshold $\tau_c$ (usually 6 to 10 dB) higher than the other signals, the receiver will preferentially decode the signal with higher power and ignore or discard the weaker signal. However, when the signal power difference is small, the receiver may switch between the two signals continuously and fail to decode either signal. This model uses the capture effect threshold $\tau_c = 6\,\text{dB}$, which is consistent with the mainstream standard of related research [4].

This model models the demodulator of the LoRaWAN gateway in detail, which is not covered by most existing simulators. In particular, this study focuses on the SX1301 LoRaWAN gateway chip produced by Semtech [5]. The SX1301 chip is capable of detecting the preambles corresponding to all SFs on 8 channels simultaneously, with a total of 48 possible combinations. However, the architecture of this chip limits the number of packets it can demodulate simultaneously to no more than 8. This is due to the design of the SX1301, which separates the preamble detection and acquisition tasks from the demodulation process. This model also considers the locking behavior of the LoRa demodulator, that is, when the demodulator detects a certain number of $\tau_p$ of preambles of the signal, it will lock on to this signal. Even if a strong signal with a power exceeding the threshold $\tau_c$ arrives later, the demodulator will not switch to demodulate the strong signal, resulting in both signals being unable to be demodulated correctly. Based on the physical experiments conducted by Rahmadhani et al. [6], this model adopts $\tau_c = 4$.

In view of the high cost of conducting physical experiments on LoRaWAN, this part of the work developed an open source simulator LoRaWANSim. LoRaWANSim can not only simulate the above-mentioned LoRa link behavior, but also implement the core functions of the LoRaWAN protocol. LoRaWANSim supports uplink and downlink communications, and implements a gateway with multi-channel multi-demodulators and a terminal node with a receiving window. The code of LoRaWANSim can be found at https://github.com/wddd1121/LoRaWANSim.

## 2.2 Problem Representation

This section aims to comprehensively consider the behavior of the LoRa link and the characteristics of the LoRaWAN standard, and to improve the overall performance of the LoRaWAN network by optimizing the parameter allocation strategy for uplink and downlink. Based on the LoRaWAN network model constructed in the previous section, the target problem of this section can be expressed as:

$$\max_{\{USF_i, P_i, DSF_i, w_i\}} \frac{1}{N} \sum_{i=i}^{N} PDR_i \qquad (4)$$

$$\text{s.t. } (4.a): RSSI_i \geq \tau_s,$$
$$(4.b): SINR_i \geq \tau_d,$$
$$(4.c): RSSI_i \geq 4 \max_{j \in \mathcal{D}_{s_i} \setminus \{i\}} \mathbb{I}_t(i,j) \mathbb{I}_c(i,j) RSSI_j,$$
$$(4.d): t_i = \arg \min_{j \in \mathcal{D}_{s_i}} \mathbb{I}_t(i,j) \mathbb{I}_c(i,j) (t_j + 4TOA_{s_i}),$$
$$(4.e): \mathcal{M} < 8.$$

Where $PDR_i$ is the $PDR$ of the terminal node $i$. The objective problem solves the optimal uplink and downlink parameter allocation strategy to maximize the average $PDR$ of all terminal nodes. Constraints (4.a) to (4.e) are the conditions that need to be met for the message to be successfully demodulated by the receiver. Constraints (4.a) and (4.b) ensure that the $RSSI$ and $SINR$ of the terminal node $i$ meet the gateway's receiving sensitivity threshold and demodulation threshold respectively. Constraint (4.c) indicates that the $RSSI$ of the terminal node $i$ must be more than four times the $RSSI$ of the conflicting signal, that is, 6dB, to meet the capture effect that occurs when the gateway captures multiple LoRa signals. Constraint (4.d) reflects the requirement that the end node must start transmitting its message before other interfering devices lock the demodulator, where $t_i$ represents the transmission time of the end node $i$, and $TOA_{s_i}$ is the time required for $SF = s_i$ to transmit a preamble. Constraint (4.e) ensures that there are available demodulators at the gateway, where $\mathcal{M}$ represents the number of occupied demodulators. This study focuses on the SX1301 chip, which has 8 demodulators.

## 2.3 Markov Decision Process Representation

This section models the resource allocation and decision-making process of the LoRaWAN network as a Markov Decision Process (MDP). MDP is a classic formal expression of sequential decision making, specifically used to model dynamic systems with randomness and decision-making. A standard MDP consists of a five-tuple $\langle \mathcal{S}, \mathcal{A}, \mathcal{P}, \mathcal{R}, \gamma \rangle$, where the state space $\mathcal{S}$ contains all the states that the system may be in. The action space $\mathcal{A}$ includes all the actions that the agent may perform in each state. The state transition probability function $\mathcal{P}(s'|s,a)$ describes the probability that the agent transitions from one state to another under a specific action. The reward function $\mathcal{R}$ represents the immediate reward obtained after performing an action. The discount factor $\gamma$ is used to weigh the relationship between immediate rewards and future rewards.

(1) State Space and Action Space
Currently, most existing research focuses on optimizing two main parameters of the uplink: spreading factor and transmit power. Increasing the spreading factor can enhance the communication range of the device, but at the same time it will reduce the data transmission rate. In addition, since LoRaWAN nodes use the ALOHA protocol for transmission, a higher spreading factor increases the possibility of transmission collisions. For battery-powered devices, increasing the transmit power will significantly shorten the battery life and may increase signal interference, especially in densely deployed environments.

In addition to the uplink spreading factor and transmit power, downlink-related parameters also have an important impact on network performance, because the downlink is responsible for transmitting parameter adjustment instructions. Since the gateway usually has only one downlink and there may be multiple uplinks, when multiple devices need to receive downlink data at the same time, using a larger spreading factor may result in insufficient downlink time and thus lost data packets. In LoRaWAN, the terminal node opens two downlink receive windows after sending uplink data. A smaller receive window may make it difficult for the gateway to send data to the terminal device within the appropriate time, while a larger receive window will affect the transmission of uplink data packets and increase energy consumption. Therefore, in the design and deployment of the LoRaWAN network, it is crucial to reasonably allocate and balance the above parameters to improve network efficiency.

Therefore, unlike Bor et al. [7] who assumed that downlink traffic is costless or sent immediately, the state space and action space in this study consider both uplink and downlink parameters. The state of the terminal node at time step is represented by a five-dimensional vector as follows:

$$s_t^i = \{d, USF, P, DSF, w\}. \tag{5}$$

Where $d$ is the distance between the end node and the gateway. $USF$ and $P$ are the SF and transmit power used by the end node in uplink communication. $DSF$ and $w$ are the SF and receive window size of the end node used by the gateway when initiating downlink communication to the end node. $\mathcal{S}$ consists of the states of all end nodes in the LoRaWAN network, expressed as $\mathcal{S} = \{s_t^i | i = 1, 2, \cdots, N\}$. After each uplink transmission, the end node starts two consecutive receive windows to receive downlink data from the gateway [8]. The LoRaWAN standard divides these windows into $RX_1$ and $RX_2$. Specifically, $RX_1$ opens one second after the end node completes the uplink transmission. If the terminal node fails to receive correct downlink data within $RX_1$, $RX_2$ will be opened two seconds after the end of the uplink transmission. If the terminal node receives downlink data within $RX_1$, $RX_2$ will not be opened.

In the LoRaWAN network, the actions taken by all terminal nodes can be composed of $\mathcal{A}$, expressed as $\mathcal{A} = \{a_t^i | i = 1, 2, \cdots, N\}$, where $a_t^i$ represents the action of terminal node $i$ at time step $t$. The actions are represented by a four-dimensional vector:

$$a_t^i = \{USF, P, DSF, w\}. \tag{6}$$

In order to correspond to $s_t^i$, $a_t^i$ uses the same notation as $s_t^i$. Here, $USF$ refers to the spreading factor that the terminal node will use in the next uplink. $DSF$, $P$ and $w$ are the spreading factor, transmit power and receive window size that the gateway will use in the next downlink, respectively.

(2) Reward Function

The reward function is a key component for evaluating the behavior of an agent. It defines a scalar reward value for each action of the agent in the environment. This function provides immediate feedback on the quality of the agent's behavior, thereby guiding its future decisions. The reward function proposed in this paper takes into account multiple key factors, including immediate PDR and cumulative PDR. The specific expression of the reward function $\mathcal{R}$ is as follows:

$$\mathcal{R}(s_t^i, a_t^i) = -2\log_2\left(1 - (1-\lambda)\frac{n_{t+1}^{succ} - n_t^{succ}}{n_{t+1}^{sent} - n_t^{sent}} - \lambda \operatorname{H}(s_t^i)\right). \tag{7}$$

Where $n_t^{sent}$ and $n_t^{succ}$ represent the total number of packets sent and the total number of packets successfully received by the terminal node $i$ at time step $t$, respectively. The ratio $(n_{t+1}^{succ} - n_t^{succ})/(n_{t+1}^{sent} - n_t^{sent})$ represents the instantaneous PDR from time step $t$ to $t+1$, providing the agent with direct feedback on its current actions, helping it to quickly adapt to changes in the network environment. The function $\operatorname{H}(s_t^i)$ represents the cumulative PDR under given parameters, which is used to evaluate long-term performance, thereby reducing the impact of short-term network fluctuations. The trade-off factor $\lambda$ is used to balance the impact of immediate and cumulative PDR, ensuring that the reward function takes into account both the current network performance and the stability of historical data. The application of negative logarithmic transformation increases the penalty for actions with low success rates, aiming to improve the training efficiency of the entire network.

# 3 LoRaWAN Resource Allocation and Decision-making Method

In order to solve the problems described in the previous section, this section proposes the HEAT algorithm to improve the overall performance of the LoRaWAN network. The HEAT algorithm uniformly considers the uplink spreading factor and transmit power that are widely concerned in related works under the LoRaWAN standard, as well as the downlink spreading factor and receive window opening size that are not considered in most related works. In the following section, this article will introduce the details of the HEAT algorithm in detail.

## 3.1 Network Model

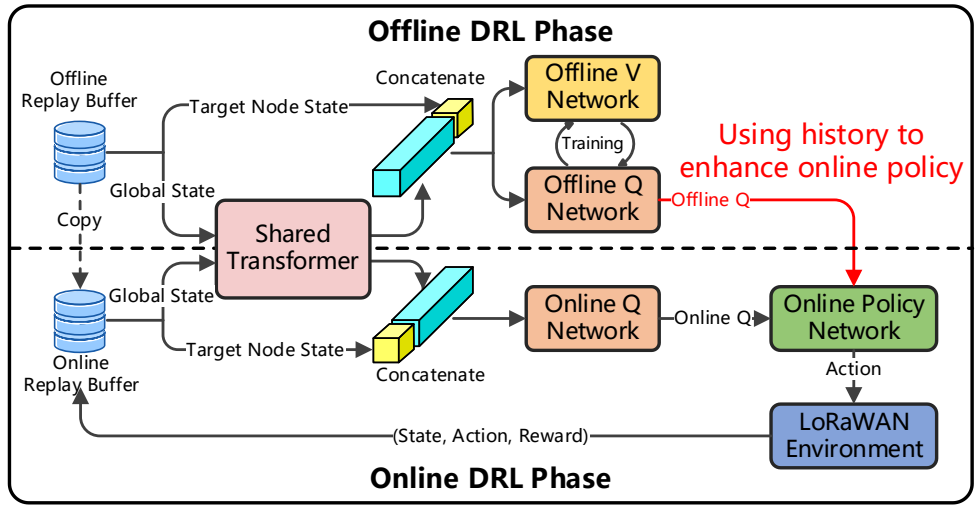

Fig.1. Model structure of HEAT algorithm

The model framework of the HEAT algorithm is shown in Fig.1. The HEAT algorithm uses the currently popular Actor-Critic architecture as the basis of the algorithm to estimate the optimal value function and policy function. In addition, the HEAT algorithm combines the advantages of offline reinforcement learning and online reinforcement learning, and makes full use of historical data to enhance model performance. In the offline reinforcement learning stage, LoRaWAN historical data is used to train the offline optimal value estimation model. In the online reinforcement learning stage, the online policy model interacts with the environment and performs online learning, and the learning process is enhanced by the offline optimal value estimation model.

The HEAT algorithm uses a shared Transformer to extract a global feature representation, expressed as $U_\vartheta$, where $\vartheta$ is a model parameter. The shared Transformer uses the encoder part of the Transformer to encode the global state and provide a unified self-correlation feature within the global state for multiple value networks and policy networks. Specifically, this paper regards all terminal nodes in the LoRaWAN network as sequence data, each terminal node as an element in the sequence data, and the state of the terminal node $s$ as an element feature, thereby using the Transformer encoder to encode all terminal nodes to generate the self-correlation feature $\mathcal{G}$ of the global state.

All value networks and policy networks in HEAT share $U_\vartheta$, both in offline and online reinforcement learning stages, which significantly reduces the number of network parameters. After obtaining $\mathcal{G}$, HEAT combines $\mathcal{G}$ with the original input of these networks (i.e., the target node state) as a new input. $\mathcal{G}$ provides a relational view of all nodes in the network, not just the local information of the target node. By adding $\mathcal{G}$, the model can understand the association information of the target node with other nodes in the entire network, thereby helping the model to make more accurate decisions. In addition, since the Transformer does not use the traditional recursive structure, it introduces position encoding to maintain the position information of elements in the input sequence. However, in the LoRaWAN wireless network environment, the positional relationship of each node in the network in the Transformer input sequence does not affect the interference of wireless signals between nodes and resource competition. Therefore, the proposed shared Transformer removes the original position encoding module to ensure that the model can focus on the relationship and mutual influence between nodes.

In the HEAT algorithm, the structures of the Actor and Critic networks are shown in Fig.2 and Fig.3. Unlike the policy networks designed in other related works, a single Actor is used in this paper to generate action probability distributions for all parameters of the LoRaWAN node. Each controllable parameter is outputted by the corresponding branch decision maker, which is concatenated to form the action probability distribution matrix of the target node. These branch decision makers have the same network structure. The policy behaviors of multiple controllable parameters share the feature extraction part, but multiple separate branch decision makers are used to calculate the decision action, which is conducive to the agent learning corresponding strategies with its own characteristics for controllable parameters with different characteristics.

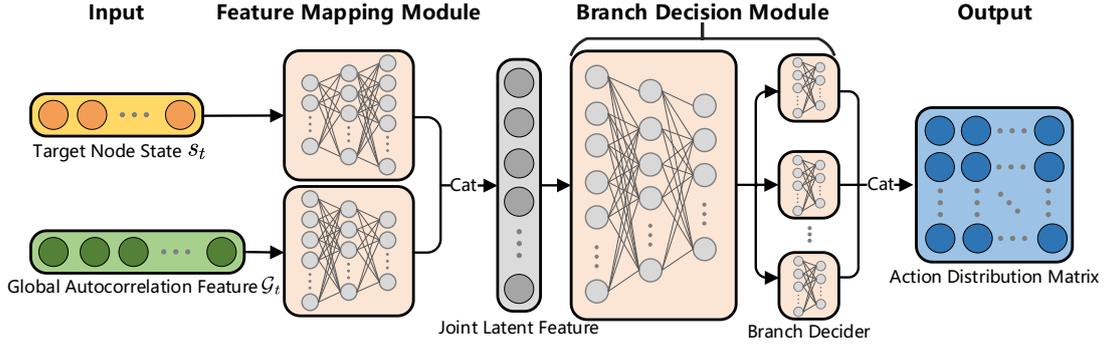

Fig.2 Actor network

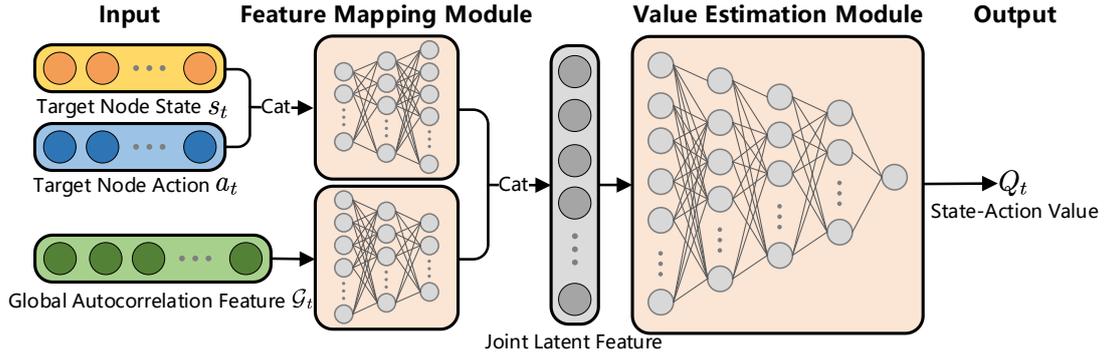

Fig.3 Critic network

The input layers of the Actor network and the Critic network are very similar. The Actor network takes the target node state $s_t$ and the global autocorrelation feature $\mathcal{G}_t$ at time $s$ as input to estimate the optimal decision distribution of the target node relative to the current network environment. The Critic network takes the target node state $s_t$, the target node action $a_t$ and the global autocorrelation feature $\mathcal{G}_t$ at time $s$ as input to estimate the value of the action taken by the target node relative to the current network environment. The global autocorrelation feature can help the model understand the association between nodes, while the local state can reflect the immediate specific situation of the specific node. By combining local and global information, the model can more specifically understand the external decision environment faced by the target node, thereby improving the accuracy of the decision.

The feature mapping modules of the Actor network and the Critic network are consistent, both of which are composed of a fully connected network and a Leaky ReLU activation function. Due to the large differences in the dimensions of the target node state $s_t$, the target node action $a_t$ and the global autocorrelation feature $\mathcal{G}_t$, the HEAT algorithm uses a different feature mapping method. For the $s_t$ and $a_t$ of the LoRaWAN node, a gradually increasing fully connected network size is used, which can map the network input to a high-dimensional space, which is conducive to reducing the coupling between each dimension of the network input. For the global autocorrelation feature $\mathcal{G}_t$, a gradually decreasing fully connected network size is used, which can reduce the feature dimension and is conducive to extracting effective features. After the two inputs are processed through independent paths, they are spliced to obtain joint hidden features. For the Actor network, after the joint hidden features are processed by multiple branch decision modules, the output layer will generate the action probability distribution matrix of the target node. These outputs will be used as the decision results of the Actor to guide the behavior of the target node in the environment. The target node obtains the specific action for the next step through sampling.

For the Critic network, after obtaining the joint latent features, the Critic network uses the value estimation module to evaluate the value of the current state-action pair and provide guidance for the training of the Actor network. It is worth mentioning that the HEAT algorithm adopts the Clipped Double Q-learning technique [9]. This technique refers to maintaining two Critic networks with the same structure in the algorithm and using these two Critic networks to estimate the value of the state-action pair at the same time to obtain $Q_1$ and $Q_2$, and finally taking $\min\{Q_1, Q_2\}$ as the final estimate of the value of the state-action pair. This technique can reduce the Critic's overestimation of the value of the state-action pair during network updates, allowing the Actor to learn the optimal strategy more accurately.

## 3.2 Offline Reinforcement Learning Method Based on In-sample Learning

For online reinforcement learning, it is usually necessary to rely on a real-time online experience replay cache to store data such as states, actions, rewards, etc. collected during the interaction between the agent and the environment to infer the dynamic laws of the environment. However, it takes time for the agent to interact with the environment, which may take a long time to obtain an effective strategy, especially in the case of sparse rewards. In addition, during the exploration process, the agent may make unreasonable decisions without sufficient experience. However, every interaction with the environment may be costly, especially in real-world applications, where some interactions may cost a lot of time or money, or even bring irreversible consequences. In the LoRaWAN network environment, since terminal nodes usually transmit data at a lower frequency, the experience replay cache takes a long time to collect enough experience data. In addition, if the model trained with an experience replay cache with insufficient data is used in the process of reinforcement online learning, the LoRaWAN network may experience severe fluctuations, or fall into a quasi-absorption state that is difficult to escape in a larger-scale network.

Offline reinforcement learning is a method in which an agent no longer relies on real-time interaction with the environment during training, but instead learns from pre-collected data. These data sets are collected from other behavioral strategies or expert demonstrations. For LoRaWAN networks, offline reinforcement learning methods can use a large amount of historical data derived from other strategies for training, such as the ADR algorithm and the random uniform strategy, so they are significantly better than online reinforcement learning methods in terms of data utilization efficiency. In addition, offline reinforcement learning does not require interaction with the LoRaWAN environment, which significantly reduces the adverse impact that the agent may have on the network during the exploration process.

Implicit Q-Learning (IQL) is a method for offline reinforcement learning in an offline training dataset [10]. It belongs to the in-sample learning paradigm and can be derived from the implicit value regularization (IVR) framework [11]. IVR is a value function regularization method used in reinforcement learning. It aims to improve the generalization ability of the model and prevent overfitting by implicitly constraining the value function, thereby helping the reinforcement learning model to converge more stably in complex environments. The core idea of IVR is to impose regularization constraints on the learning process in an indirect way without explicitly calculating or restricting the specific form of the value function to reduce the risk of overfitting of the model. In IQL, the model is trained by minimizing the loss function on the offline training data, and the goal is to find the parameter configuration that minimizes the error of the offline training data. Since the IQL method runs in an offline reinforcement learning manner, IQL can use limited offline data to provide a more satisfactory strategy for the LoRaWAN network, reduce the fluctuation of the LoRaWAN network and improve the learning efficiency. Therefore, in the offline reinforcement learning stage, HEAT uses the IQL method to estimate the optimal value function in the offline training data set.

Algorithm 1 describes the offline reinforcement learning process based on in-sample learning in the HEAT algorithm. The offline deep reinforcement learning stage uses the pre-collected offline experience replay cache $\mathcal{B}_{off}$ to solve the state value function $V_\psi^\mu(s)$ and the state-action pair value function $Q_\phi^\mu(s,a)$. $\psi$ and $\phi$ are the parameters of the function, respectively, and $\mu$ represents the optimal policy in $\mathcal{B}_{off}$. Among them, $V_\psi^\mu(s)$ and $Q_\phi^\mu(s,a)$ respectively estimate the optimal state value function and the optimal state-action pair value function corresponding to $\mu$. HEAT only uses samples within $\mathcal{B}_{off}$ to learn $Q_\phi^\mu(s,a)$, thereby avoiding exploring actions outside the offline experience replay cache. It is worth noting that in the IQL method, there is no need to explicitly determine the specific form of $\mu$, but to directly solve its associated value function. The process of solving $V_\psi^\mu(s)$ and $Q_\phi^\mu(s,a)$ is as follows:

$$\underset{V_\psi^\mu}{\operatorname{argmin}} \mathbb{E}_{(s,a) \sim \mathcal{B}_{off}} \left[ L_2^\rho \left( Q_\phi^\mu(s,a) - V_\psi^\mu(s) \right) \right]. \tag{8}$$

$$\underset{Q_\phi^\mu}{\operatorname{argmin}} \mathbb{E}_{(s,a,r,s') \sim \mathcal{B}_{off}} \left[ (r + \gamma V_\psi^\mu(s') - Q_\phi^\mu(s,a))^2 \right]. \tag{9}$$

**Algorithm 1: In-sample Offline Reinforcement Learning Algorithm**

**Input:** offline Q network $Q_{\phi_1}$, $Q_{\phi_2}$, offline V network $V_{\psi_1}$, $V_{\psi_1}$, shared transformer $U_\vartheta$, offline experience replay buffer $\mathcal{B}_{off}$

**Outout:** Updated network parameters $\phi_1$, $\phi_2$, $\psi_1$, $\psi_2$, and $\vartheta$

1: Random Initialize $\phi_1$, $\phi_2$, $\psi_1$, $\psi_2$, and $\vartheta$
2: Set the target parameters to be equal to the main parameters:
$$\phi_i^{\text{targ}} \longleftarrow \phi_i \qquad \text{for } i = 1, 2$$
$$\psi_i^{\text{trag}} \longleftarrow \psi_i \qquad \text{for } i = 1, 2$$
3: **Repeat**
4:     Randomly sample a batch of transitions from $\mathcal{B}_{off}$:
$$B = \{\{(s^i, a^i, r^i, s'^i) \mid i = 1, 2, \cdots, N\}\}$$
5:     Calculate global autocorrelation features $\mathcal{G} \longleftarrow U_\vartheta(B)$
6:     The formula for the temporal-difference target value of offline V network:
$$Y_q^{\text{targ}}(r, s') = r + \gamma \min_{i=1,2} V_{\psi_i^{\text{targ}}}(s', \mathcal{G})$$
7:     The target value formula of the offline Q network:

$$Y_v^{\text{targ}}(s,a) = \min_{i=1,2} Q_{\phi_i^{\text{targ}}}(s,a,\mathcal{G})$$

8:     Calculate the temporal-difference loss of offline Q network:

$$J_{\phi_i} \longleftarrow \frac{1}{|B|} \sum_{(s,a,r,s') \in B} (Y_q^{\text{targ}}(r,s') - Q_{\phi_i}(s,a))^2 \quad \text{for } i=1,2$$

9:     Calculate the loss of offline V network:

$$J_{\psi_i} \longleftarrow \frac{1}{|B|} \sum_{(s,a) \in B} L_2^\rho (Y_v^{\text{targ}}(s,a) - V_{\psi_i}(s,\mathcal{G})) \quad \text{for } i=1,2$$

10:    Calculate gradient of $\phi_1$, $\phi_2$, $\psi_1$, $\psi_2$, and $\vartheta$ (use gradient accumulation):

$$\nabla_{\phi_i}, \nabla_\vartheta \longleftarrow \nabla J_{\phi_i} \quad \text{for } i=1,2$$
$$\nabla_{\psi_i}, \nabla_\vartheta \longleftarrow \nabla J_{\psi_i} \quad \text{for } i=1,2$$

11:    Use a gradient optimizer to update network parameters, such as Adam:

$$\phi_1, \phi_2, \psi_1, \psi_2, \vartheta \longleftarrow \text{Adam}(\nabla_{\phi_1}, \nabla_{\phi_2}, \nabla_{\psi_1}, \nabla_{\psi_2}, \nabla_\vartheta)$$

12:    Update target parameters:

$$\phi_i^{\text{targ}} \longleftarrow \phi_i \quad \text{for } i=1,2$$
$$\psi_i^{\text{trag}} \longleftarrow \psi_i \quad \text{for } i=1,2$$

13:   **Until** HEAT algorithm convergence

---

The expectation of $V_\psi^\mu(s)$ is defined as the solution to the asymmetric least squares problem $L_2^\rho(x) = |\rho - \mathbb{1}(x<0)|x^2$. Where $\rho$ is a weighting factor to control the regularization strength. When $\rho > 0.5$, this asymmetric loss function reduces the weight of state-action pair values that are smaller than the state value, and gives more weight to larger state-action pair values. $V_\psi^\mu(s)$ can be viewed as an estimate of the maximum state-action pair value supported by actions in $\mathcal{B}_{off}$, which can then be used to update $Q_\phi^\mu(s,a)$ using the mean squared error loss. It should be noted that these losses do not use any explicit strategy, but simply use the actions in the dataset to achieve two goals.

## 3.3 Online Reinforcement Learning Method Based on Historical Enhancement

Since offline reinforcement learning can extract knowledge from historical data and does not require interaction with the online environment, while online reinforcement learning can learn the personalized needs of the intelligent agent and adapt to changes in the environment through interaction with the online environment, more and more research works have combined offline reinforcement learning with online reinforcement learning to improve learning efficiency and reduce training costs [12]. Specifically, it can be divided into an offline pre-training stage and an online fine-tuning stage. In the offline pre-training stage, a baseline model that performs well in the environment is generated through offline reinforcement learning using the pre-historical data. In the online fine-tuning stage, the baseline model is deployed to the environment, and the baseline model is fine-tuned using real-time data and environmental feedback, so that the baseline model can adapt to changes and uncertainties in the actual environment. However, this method optimizes the same model in the offline training stage and the online training stage, which makes it impossible to adjust the constraint strength of the learned offline optimal policy on the online training process. In addition, since the same model is shared, the online learned policy may overwrite the offline policy, resulting in the problem of forgetting historical experience.

Inspired by the separate training in the Offline-Boosted Actor-Critic (OBAC) algorithm [13], the offline learning phase and the online learning phase of the HEAT algorithm adopt parallel separate training. The basic process is as follows: In the offline training phase, HEAT uses the previous historical data to generate a model $\text{Model}_{\text{off}}$ that performs well in the environment through offline reinforcement learning. In the online training phase, another model $\text{Model}_{\text{on}}$ is deployed to the environment for online reinforcement learning, and $\text{Model}_{\text{off}}$ is used to enhance this process. In addition, during the online training phase, the offline training of $\text{Model}_{\text{off}}$ is still in progress. Therefore, the training process of $\text{Model}_{\text{off}}$ and $\text{Model}_{\text{on}}$ can be executed in parallel, one focusing on mining offline data and the other focusing on exploring the online environment, so that both sample utilization and training efficiency are improved. $\text{Model}_{\text{on}}$ can make faster adjustments based on the knowledge of $\text{Model}_{\text{off}}$, thereby enhancing the adaptability of the entire system. In addition, because $\text{Model}_{\text{off}}$ and $\text{Model}_{\text{on}}$ are trained separately, the constraint strength of $\text{Model}_{\text{off}}$ on the training process of $\text{Model}_{\text{on}}$ can be easily adjusted. The training process of $\text{Model}_{\text{off}}$ is not affected by $\text{Model}_{\text{on}}$, so the strategy learned by $\text{Model}_{\text{on}}$ will not overwrite $\text{Model}_{\text{off}}$, and the problem of forgetting historical experience will not occur.

In the online deep reinforcement learning phase, HEAT uses a dynamically updated online experience replay cache $\mathcal{B}_{on}$ and a state-action pair value function $Q_\varphi^\pi(s,a)$ to optimize the online learning policy $\pi_\theta(a|s)$, where $\theta$ and $\varphi$ are the parameters of their respective functions. HEAT adopts the popular Actor-Critic framework. In this architecture, the agent selects actions according to the policy $\pi_\theta(a|s)$ and evaluates the selected actions through $Q_\varphi^\pi(s,a)$. Subsequently, the policy $\pi_\theta$ is updated based on the evaluation results to improve its decision-making performance. At the same time, $Q_\varphi^\pi$ is also updated to improve the evaluation accuracy of future state-action pairs. New samples $(s,a,s',r)$ collected during the interaction between the agent and the environment are incorporated into $\mathcal{B}_{on}$. HEAT optimizes this function by minimizing the mean square error between $Q_\varphi^\pi(s,a)$ and its temporal difference target. The specific objective function is:

$$\underset{Q_\varphi^\pi}{\arg\min}\, \mathbb{E}_{(s,a,r,s')\sim \mathcal{B}_{on}}\left[\left(r+\gamma \max_{q\in A} Q_\varphi^\pi(s',q)-Q_\varphi^\pi(s,a)\right)^2\right]. \tag{10}$$

HEAT algorithm proposes the history enhancement technique to improve the training efficiency of the model. The history enhancement technique is mainly reflected in the optimization process of the reward function $\mathcal{R}$ and the strategy. As described in Section 4.1.3, the global history reward $H(USF_t^i, P_t^i, DSF_t^i, w_t^i)$ and the local history reward $(n_{t+1}^{succ} - n_t^{succ})/(n_{t+1}^{sent} - n_t^{sent})$ in $\mathcal{R}$ can provide nodes with global experience from the online learning history and immediate value feedback of the current action, which is conducive to providing guidance information for decision-making. Next, this paper will elaborate on the optimization process of the strategy. After offline reinforcement learning, the value function $V_\psi^\mu(s)$ and the state-action pair value function $Q_\phi^\mu(s,a)$ are solved, which represent the estimate of the optimal value function in the offline replay cache. For these functions, it is natural to think of using them to enhance the optimization process of the policy $\pi_\theta(a|s)$. This paper improves the policy optimization step in the OBAC algorithm and introduces value margin optimization to improve the optimization efficiency of $\pi_\theta(a|s)$:

$$\operatorname*{argmax}_{\pi_\theta} \mathbb{E}_{s \sim \mathcal{B}_{on}}[\mathbb{E}_{a \sim \pi_\theta}[Lesson_{on}(s,a)] + \mathbb{E}_{a \sim \mathcal{B}_{on}}[Lesson_{off}(s,a)]],$$

$$Lesson_{on}(s,a) = \alpha \cdot \operatorname{Ent}(s,a,\pi_\theta) + Q_\varphi^\pi(s,a) \cdot \log \pi_\theta(a|s), \qquad (11)$$

$$Lesson_{off}(s,a) = \beta \cdot \operatorname{Mar}(Q_\phi^\mu(s,a) - Q_\varphi^\pi(s,a)) \cdot \log \pi_\theta(a|s).$$

Where $Lesson_{on}$ and $Lesson_{off}$ are the online lesson term and the historical lesson term, respectively. $Lesson_{on}$ is the lesson learned by the agent in the process of interacting with the environment, with an entropy regularization term. $Lesson_{off}$ is the lesson learned by the agent throughout history. Among them, $\operatorname{Ent}(s,a,\pi_\theta)$ is the logarithm of the probability of the strategy $\pi_\theta$ taking all actions in state $s$ except action $a$, which helps prevent the strategy from converging to a suboptimal local solution too early to enhance exploration capabilities. $\operatorname{Mar}(x) = \mathbb{1}(x>0) \cdot x$ is used to select actions when the offline state-action pair value is greater than the online state-action pair value, and calculate the value margin. Subsequently, HEAT uses the obtained value margin to weight different actions, and realizes the optimization process of enhancing the online policy $\pi_\theta$ with the offline optimal policy. The parameters $\alpha$ and $\beta$ are trade-off factors used to adjust the contribution of each item.

To speed up the training process of the overall model, in the online reinforcement phase, the HEAT algorithm merges the offline replay cache $\mathcal{B}_{off}$ with the online replay cache $\mathcal{B}_{on}$ to form a hybrid replay cache $\mathcal{B}$. Since the neural network parameters are randomly initialized at the beginning of training, unstable outputs or gradients may occur. To this end, the HEAT algorithm adopts a $\gamma$ incremental strategy to reduce the instability in the early stage of training and avoid gradient explosion or excessive Q value fluctuations caused by large discount factors. Algorithm 2 describes the offline reinforcement learning process based on in-sample learning in the HEAT algorithm.

**Algorithm 2: History-enhanced Online Reinforcement Learning Algorithm**

**Input:** online Q network $Q_{\varphi_1}$, $Q_{\varphi_2}$, online policy network $\pi_\theta$, shared transformer $U_\vartheta$, online experience replay buffer $\mathcal{B}_{on}$, current network time $t$

**Outout:** Updated network parameters $\varphi_1$, $\varphi_2$, $\theta$, and $\vartheta$

1: Set the target parameters to be equal to the main parameters:
$$\varphi_i^{\text{targ}} \longleftarrow \varphi_i \qquad \text{for } i = 1, 2$$

2: **Repeat**

3:   The network server receives the uplink message from node $i$

4:   Policy network generates decision actions $a_t^i \longleftarrow \pi_\theta(s_t^i)$

5:   The network server sends $a_t^i$ to the node $i$

6:      $\mathcal{B}_{on} \longleftarrow \mathcal{B}_{on} \cup \{s_t^i, a_t^i, r_t^i, s'^i_{t-1}\}$

7:      Randomly sample a batch of transitions from $\mathcal{B}_{on}$:

$$B = \{\{(s^i, a^i, r^i, s'^i) \mid i = 1, 2, \cdots, N\}\}$$

8:      Calculate global autocorrelation features $\mathcal{G} \longleftarrow U_\vartheta(B)$

9:      The formula for the temporal-difference target value of online Q network:

$$Y_q^{\text{targ}}(r, s') = r + \gamma \min_{i=1,2} \max_{q \in A} Q_{\varphi_i^{\text{targ}}}^{\pi_{\text{targ}}}(s', q, \mathcal{G})$$

10:     The formulas for online lesson item and historical lesson item:

$$Lesson_{on}(s, a) = \alpha \operatorname{Ent}(s, a, \pi_\theta) + (\min_{i=1,2} Q_{\varphi_i}^\pi(s, a)) \log \pi_\theta(a \mid s)$$

$$Lesson_{off}(s, a) = \beta \operatorname{Mar}(\min_{i=1,2} Q_{\phi_i}^\mu(s, a) - \min_{i=1,2} Q_{\varphi_i}^\pi(s, a)) \log \pi_\theta(a \mid s)$$

11:     Calculate the temporal-difference loss of offline Q network:

$$J_{\varphi_i} \longleftarrow \frac{1}{|B|} \sum_{(s,a,r,s') \in B} (Y_q^{\text{targ}}(r, s') - Q_{\varphi_i}(s, a))^2 \quad \text{for } i = 1, 2$$

12:     Calculate the loss of policy network:

$$J_\pi \longleftarrow \frac{1}{|B|} \sum_{(s,a) \in B} (Lesson_{on}(s, \pi_\theta(s)) - Lesson_{off}(s, a))$$

13:     Calculate gradient of $\varphi_1$, $\varphi_2$, $\theta$ and $\vartheta$ (use gradient accumulation):

$$\nabla_{\varphi_i}, \nabla_\vartheta \longleftarrow \nabla J_{\varphi_i} \quad \text{for } i = 1, 2$$
$$\nabla_\theta, \nabla_\vartheta \longleftarrow \nabla J_\pi$$

14:     Use a gradient optimizer to update network parameters, such as Adam:

$$\varphi_1, \varphi_2, \theta, \vartheta \longleftarrow \operatorname{Adam}(\nabla_{\varphi_1}, \nabla_{\varphi_2}, \nabla_\theta, \nabla_\vartheta)$$

15:     Update target parameters:

$$\varphi_i^{\text{targ}} \longleftarrow \varphi_i \quad \text{for } i = 1, 2$$

16: **Until** HEAT algorithm convergence

## 4 Experimental Evaluation and Results Analysis

This section conducts experimental evaluation and results analysis of the HEAT algorithm proposed in this paper. The experiments are mainly divided into comparative experiments and ablation experiments. This paper compares HEAT with existing learning methods to prove the superiority of HEAT over existing methods. In the ablation experiment, this paper performs ablation verification on the offline part of HEAT to verify the effectiveness of the HEAT algorithm.

### 4.1 Experimental Setting

In order to demonstrate the advantages of the HEAT algorithm over other methods, this paper selected three popular algorithms of different categories for comparison under different node densities and traffic intensities: ADRx[14] is an improvement on the SARD algorithm, in which the margin_db parameter of the LoRaWAN standard is dynamically adjusted; A2C[15] is a popular deep reinforcement learning algorithm method widely used to solve decision-making problems, which adopts the Actor-Critic framework; Qmix[16] is a multi-agent reinforcement learning method that models each terminal device in the environment as an independent agent. To eliminate the randomness of the experimental results, this paper repeated the experiments of all algorithms 8 times in each case, and each experiment used a random seed of 0 to 8. All experiments were carried out under the Ubuntu 20.04 system, mainly using Python 3.9.18 and Pytorch 2.2.1 to implement the HEAT related experiments, and the CUDA version was 12.4. In terms of hardware, the GPU is RTX 4060 and the CPU is Intel(R) Xeon(R) CPU E5-2683.

This study uses Semtech's SX1278 and SX1301 chips as communication chips for terminal nodes and gateways, respectively. Among them, the gateway has 8 uplink channels and a physical demodulator. In addition, the experiment follows the CN470-510 channel plan in the LoRaWAN regional parameters. The BW of the communication link is 125 kHz, the CR is 4/5, and the carrier frequency $f$ is set to 470 kHz. In addition, it is assumed that the hybrid antenna gain $G$ is 1. In the HEAT algorithm, the offline replay buffer $\mathcal{B}_{off}$ collects data through a random uniform strategy for 25 minutes. When evaluating the performance of different algorithms, this study uses the following two performance indicators:

1. Packet Delivery Ratio (PDR): The ratio of the total number of packets successfully received by the gateway to the total number of packets sent by the end nodes。
2. Energy Efficiency Ratio (EER): The ratio of the total number of packets successfully received by the gateway to the total energy consumed by the terminal node, which reflects the transmission efficiency of the terminal node per unit energy. The energy consumption of the terminal node is estimated based on the energy model based on its air time. [15]

## 4.2 Comparative Experiment

(1) Comparative experimental results under different node densities

This section evaluates the performance of the proposed HEAT algorithm and its comparative algorithms at different node densities by analyzing PDR and EER. Fig.4 shows the changes in PDR and EER when the number of nodes $N$ increases from 128 to 768 in a LoRaWAN network with a traffic intensity $\delta$ of 2.

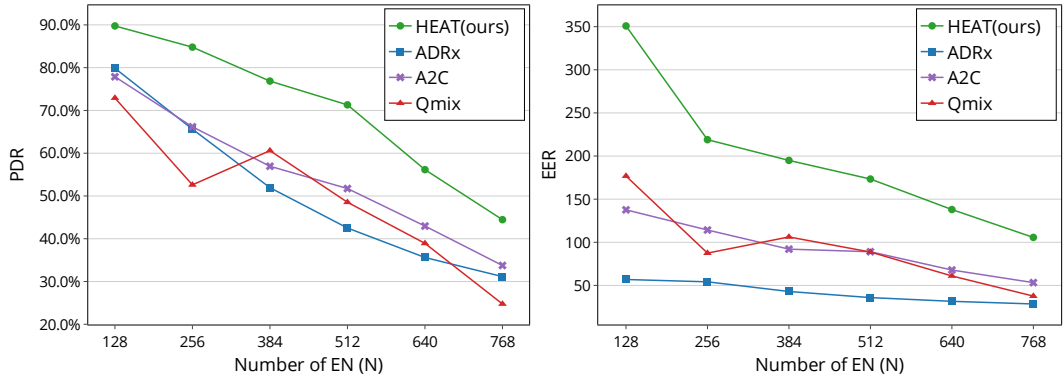

Fig.4 PDR and EER at different node densities

In general, HEAT outperforms the other three compared algorithms in PDR

performance at all node scales. Specifically, HEAT's PDR improves by an average of 18% in each scenario, and by an average of 15% over the best results of the other algorithms. HEAT performs particularly well at smaller network scales (i.e., 128 and 256 nodes), with PDRs of 90% and 85%, respectively, significantly higher than ADRx's 80% and A2C's 66%. This result shows that HEAT has a clear performance advantage under light load conditions. As the number of nodes increases to 384 and 512, the PDRs of ADRx and A2C drop significantly, while Qmix also shows large fluctuations. In contrast, HEAT's PDR decreases slightly, but the decline is limited, and it still maintains a clear advantage. When the number of nodes further increases to 640 and 768, although HEAT's success rate drops from 71% to 56% and finally to 44%, its performance is still better than other algorithms. This phenomenon shows that the HEAT algorithm still maintains significant robustness under high load conditions, and its resource allocation strategy is more effective in dealing with the network load challenges brought by the increase in the number of nodes.

The HEAT algorithm shows good EER at all node scales. Although the EER of HEAT decreases with the increase in the number of nodes, its EER value is always higher than that of other compared algorithms in the scenarios of various node scales. Specifically, the EER of HEAT increases by an average of 197% in each scenario, which is an average of 95% higher than the best results of other algorithms. HEAT enables each node to successfully send an average of 197 correctly demodulated data packets per unit energy. This performance is nearly 95% higher than the best results of the other three algorithms. In terms of EER, ADRx performs the worst, and its EER values at all node scales are significantly lower than those of other algorithms, while A2C and Qmix have similar EER performances. This finding effectively reveals that simply increasing the transmit power of ED cannot effectively improve PDR, but may lead to a decrease in PDR. The experimental results show that as the number of nodes increases, the HEAT algorithm can more effectively manage energy resources and ensure efficient transmission of data packets.

(2) Comparative experimental results under different traffic intensities

This section evaluates the performance of the proposed HEAT algorithm and its comparative algorithm under different traffic intensity scenarios. Fig.5 shows the changes in PDR and EER when the traffic intensity $\delta$ increases from 0.5 to 3 in a LoRaWAN network with 1000 terminal nodes $N$.

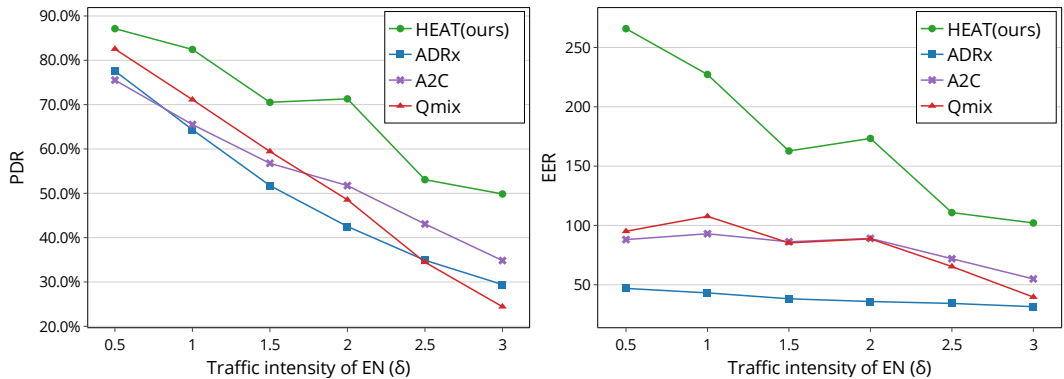

Fig.5 PDR and EER under different traffic intensities

As the traffic intensity increases, the PDR of each algorithm generally shows a downward trend. This phenomenon is mainly because the increase in the number of packets intensifies the channel competition, thereby increasing the probability of packet collision and reducing the probability of successful transmission. The PDR of HEAT increased by an average of 16% in each scenario, which is an average increase of 12% over the best results of other algorithms. At lower traffic intensities, the packet success rate of HEAT is slightly higher than that of the other three algorithms. When the traffic intensity increases to 2, HEAT shows a gradually increasing advantage over the other algorithms, and the average PDR increases to 12.6%. Even under high load conditions with traffic intensities of 2.5 and 3, although the packet success rate of HEAT drops to 53% and 50% respectively, its performance is still significantly better than that of competing algorithms. These results show that the HEAT algorithm shows stronger robustness and better network resource management efficiency in handling high traffic environments.

In terms of EER, the HEAT algorithm significantly outperforms other compared algorithms at all traffic intensity levels, especially under low to medium traffic intensities. Under various traffic intensity conditions, the PDR of HEAT increases by an average of 190%, which is an average increase of 102% over the best results of other algorithms. In particular, under the conditions of traffic intensity of 0.5 and 1, the EER of HEAT reached 265.89 and 227.25 respectively, which is much higher than the best performing Qmix, whose EER value is only 95.07 and 107.67. This data shows that in a low-traffic environment, the HEAT algorithm can more effectively utilize energy resources, thereby improving the successful transmission rate of data packets. When the traffic intensity increases from 1.5 to 3, the EER of HEAT, ADRx and A2C all show a significant decrease, accompanied by similar fluctuations. This phenomenon reflects the impact on algorithm performance as the degree of network congestion increases. Although the performance of all algorithms has declined under high traffic intensity, HEAT still shows relatively the best performance. This shows the excellent adaptability and robustness of the HEAT algorithm under different traffic intensities.

## 4.3 Ablation Experiment

This section conducts ablation verification and analysis on the proposed HEAT algorithm. In the ablation experiment, the offline reinforcement learning module of HEAT is deleted and called HEAT-online, which is compared with HEAT. This verifies the contribution of the offline reinforcement learning module to the performance of the HEAT algorithm and the advantages of combining online reinforcement learning with offline reinforcement learning.

(1) Ablation experiment results under different node densities

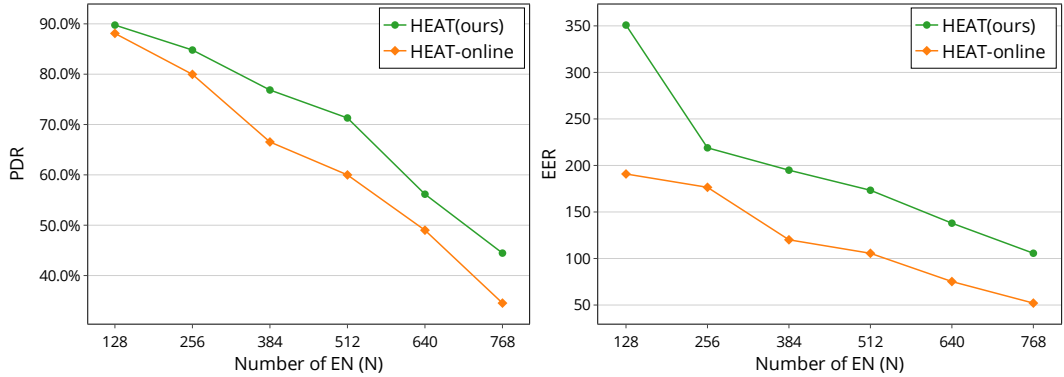

Fig.6 PDR and EER at different node densities

Fig.6 shows the changes in PDR and EER when the number of nodes $N$ increases from 128 to 768. In terms of PDR, as the number of nodes increases, both the HEAT algorithm and the HEAT-online algorithm have a downward trend in their PDR. But in general, the decline in the PDR of the HEAT algorithm is small and remains at a relatively high level, and the HEAT-online algorithm has increased by an average of 7.3%. The performance loss of the HEAT-online algorithm increases rapidly with the increase in the number of nodes, and both remain at a low level. Especially when the number of nodes is small, that is, from 128 nodes to 512 nodes, the PDR of the HEAT algorithm performs well and is relatively stable, only dropping from 90% to 71%. The PDR of the HEAT-online algorithm is very steep, dropping from 88% to 60%. In terms of EER, as the number of nodes increases, the EER of the HEAT algorithm and the HEAT-online algorithm both tend to decrease, but the EER performance of the HEAT algorithm at all node densities is better than that of the HEAT-online algorithm, which shows that the HEAT algorithm can maintain energy efficiency well at multiple node densities. Specifically, the EER of the HEAT algorithm is improved by 70.1% on average, which means that under the same energy constraint, the HEAT algorithm can send 70% more successfully demodulated messages than the HEAT-online algorithm, greatly extending the life cycle of the node.

(2) Ablation experiment results at different flow intensities

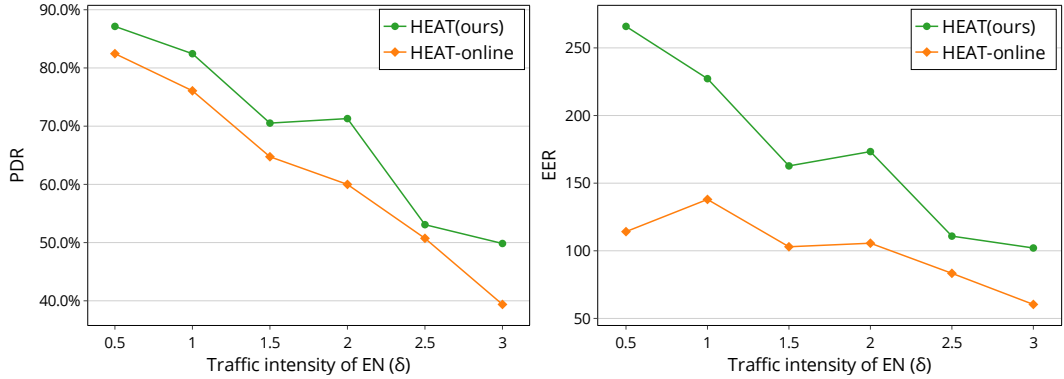

Fig.7 PDR and EER under different traffic intensities

Fig.7 shows the changes in PDR and EER when the traffic intensity $\delta$ increases from 0.5 to 3. In terms of PDR, both algorithms show a downward trend with increasing traffic intensity. However, the HEAT algorithm always has a higher PDR than HEAT-online at all traffic intensities, with an average improvement of 6.8% for the HEAT-online algorithm. Especially when there are a large number of nodes, such as traffic intensities of 2 and 3, the PDR of the HEAT algorithm is about 10% higher than that of HEAT-online. In terms of EER, HEAT shows a higher energy efficiency ratio than HEAT-online at all traffic intensities, with an average improvement of 70.3%. The difference is particularly significant at lower traffic intensities. At traffic intensities of 0.5 and 1, the EER of HEAT is improved by 132.9% and 64.7% respectively relative to HEAT-online.

In low-power wide-area networks such as LoRaWAN, as the number of nodes increases, the competition and interference faced by each node in the network also increase dramatically, resulting in the continuous expansion of the scale of the LoRaWAN network optimization problem. HEAT combines offline reinforcement learning methods with online reinforcement learning methods. The offline reinforcement learning module helps to optimize the decision-making of the algorithm before deployment, so that it can better handle different network scenarios. Therefore, for the continuous expansion of the scale of the LoRaWAN network optimization problem, this offline pre-learning ability enables HEAT to still maintain good PDR and EER, and improve the overall performance of the LoRaWAN network. The HEAT-online algorithm relies only on online reinforcement learning, which means that the strategy can only be optimized by interacting with the LoRaWAN environment. Since the HEAT-online algorithm has no prior knowledge about the LoRaWAN network environment when it is deployed, the HEAT-online algorithm may require a lot of exploration to obtain a better strategy when the scale of the optimization problem is large. However, a large amount of exploration may cause the LoRaWAN network environment to fluctuate violently, resulting in a decrease in network performance.

# 5 Conclusion

     This chapter comprehensively discusses the congestion and energy consumption problems faced by LoRaWAN networks when the node density and traffic intensity increase, from system modeling, problem statement to algorithm design and experimental verification. First, this chapter models the physical layer factors such as node deployment, signal propagation and capture effect in LoRaWAN networks, and expresses the resource allocation problem as a Markov decision process. Subsequently, the HEAT algorithm is proposed by combining offline reinforcement learning with online reinforcement learning. HEAT extracts global state autocorrelation features with the help of shared Transformer, and realizes the joint optimization of uplink and downlink parameters through offline data pre-training and online real-time fine-tuning, thereby effectively reducing network interference and improving the success rate and energy efficiency of data transmission.
     The experimental part verifies the superior performance of the HEAT algorithm under different node densities and traffic intensities through multiple groups of comparative experiments and ablation experiments. The comparative experimental results show that in each scenario, the PDR index and EER index of HEAT are better than other comparative methods; and the ablation experiment further shows that the offline reinforcement learning module has a significant promoting effect on the overall performance. Overall, the HEAT algorithm demonstrates high training efficiency and good adaptability while taking into account both network reliability and energy-saving requirements.